# LABEL-FREE ASSAYING OF TESTOSTERONE AND GROWTH HORMONES IN BLOOD USING SURFACE-ENHANCED RAMAN SPECTROSCOPY

**Annah M. Ondieki[1*], Zephania Birech[**,1], Kenneth A. Kaduki[1], Peter W. Mwangi[2], Moses Juma[1,3]**

[1] Laser Physics and Spectroscopy Research Group, Department of Physics, University of Nairobi, P.O Box 30197-00100, Nairobi, Kenya, [2] Department of Medical Physiology, University of Nairobi, P.O Box 30197-00100, Nairobi, Kenya, [3] UNESCO-UNISA Africa Chair in Nanoscience/Nanotechnology, College of Graduate Studies, University of South Africa (UNISA) South Africa, P.O Box 392 UNISA 0003, South Africa

Corresponding Authors: [*] moraa94annah@gmail.com (OA); [**] birech@uonbi.ac.ke (BZ);

## Abstract

This work reports the potential use of surface-enhanced Raman spectroscopy (SERS) in rapid, label-free assaying of testosterone (TE) and growth hormone (GH) in whole blood. Biomarker SERS spectral bands from the two hormones (TE and GH) in intentionally spiked water for injection and in male Sprague-Dawley (SD) rat's blood are reported. Abuse of the two hormones (TE and GH) singly or simultaneously is widespread and not only has prolonged side effects such as hypertension and liver failure, but their illegal use by athletes is against clean competition. Currently used highly label-dependent doping detection methods involve complex and time-consuming procedures, rendering them unsuitable for rapid analysis. In blood, the most concentration-sensitive bands (in both TE and GH), as deduced through Principal Component Analysis (PCA) and Analysis of Variance (ANOVA), were around 684 cm$^{-1}$ (assigned to C-C stretching) and 1614 cm$^{-1}$ (assigned to C-C stretching) in GH; and 786 cm$^{-1}$ (assigned to N-H wagging), 856 cm$^{-1}$ (assigned to C-C stretching), and 1490 cm$^{-1}$ (assigned to CH$_2$ bending) in TE. In addition, a characteristic variance was noted in the bands around 1510 cm$^{-1}$ (attributable to CH$_2$ stretching) in GH and 1636 cm$^{-1}$ (C-C stretching) in TE, which could be used as biomarker bands for the respective hormones in the blood. This work has shown the capability of SERS for potential hormone concentration level determination when concentration-sensitive or biomarker bands are employed. This discovery opens new possibilities for the use of SERS in fields such as sports science, clinical diagnostics, and biomedical research.

**Introduction**

Surface Enhanced Raman Spectroscopy (SERS) is a powerful analytical spectroscopic technique that allows the detection and characterization of molecules. It enhances the weak Raman scattering signals by utilizing, in addition to the incident, induced local fields produced by driven plasmonic oscillations in metallic nanostructures that are in proximity to the probed molecules [1]. The enhancement factor of SERS can be as high as $10^{14}$, which makes it possible to detect even single molecules [2]. In addition, SERS provides molecular-specific information that can be used to identify, characterize, and quantify complex molecular mixtures. As a result, SERS has been used to identify and characterize biomolecules such as proteins [3], nucleic acids, and carbohydrates [4]; and environmental pollutants including heavy metals [5] and pesticides [6]. Its ability to detect molecules at low concentrations makes it a valuable tool for environmental and medical applications [7]. SERS biomedical applications have recently grown in popularity, with distinct types of bacteria being detected and identified based on their unique SERS spectra [8]. This makes it possible for rapid and reliable identification of bacterial diseases [8]. Biomarkers such as diabetes type 2 biomarkers [9] and cancer biomarkers [10] have been as well detected using the SERS technique.

Testosterone (TE) and growth hormone (GH) are two hormones that play crucial roles in the regulation of many physiological processes in humans, including muscle growth, development and power, bone density, and sexual function [11,12]. In addition, these hormones are of interest to the sports world, where they are often abused singly or simultaneously as performance-enhancing drugs [13,14], often without knowledge of their prolonged side effects, such as hypertension and liver failure [15]. This is due to the drive to win competitions and improve athletes' visual appearance [16]. These dopants harm users and their use is against clean competition [17]. Therefore, detecting these hormones in the blood is of paramount importance in both the medical and sports-related fields. The most common methods for determining the levels of these hormones in the blood are currently based on immunoassays, such as serum and enzyme-linked immunoassay (ELISA) [12,17], and mass spectrometry, such as liquid chromatography-mass spectrometry (LC/MS) [18] and ultra-high-performance liquid chromatography–high-resolution mass spectrometry (UHPLC/HRMS) [19]. These methods provide accurate results but are often limited in cost, sensitivity, and selectivity, require complex sample preparation, and cannot simultaneously assay both GH and TE [20]. Therefore, there is a great need to develop novel methods for the detection of these hormones.

Spectroscopic techniques have recently shown potential for hormone detection. For instance, Raman Spectroscopy has successfully detected female reproductive hormones in mice's blood [21,22]; Fourier Transform Infrared Spectroscopy has identified reproductive hormones in urine [23]; and SERS has identified estrogen [24], and detected TE [25,26]. Among these techniques, SERS is the most powerful. The work reported in this paper involved, first, obtaining characteristic Raman spectra of both TE and GH. Second, the existence of concentration-sensitive Raman bands in the blood for both the TE and GH was investigated. Our findings revealed that unique SERS bands that are sensitive to the levels of the two hormones in the blood exist, as well as some common hormone bands.

**Materials and Methods**

The SERS substrates (AgNPs) (150 µL) were mixed with 30 µL of the sample (pure reference solution and blood) and stirred thoroughly to obtain a homogenous mixture. 2 µl of the resulting mixture were then dropped onto aluminum-wrapped microscope glass slides (25.4 mm × 76.2 mm × 1.2 mm dimensions) and left to air dry at room temperature for sixty (60) minutes. The sample sets included pure GH (3.7 mg/ml) (GeneScience Pharmaceuticals Co., Ltd.), pure TE (250 mg/ml) (HEPIUS Pharmaceuticals, Hong Kong), and blood from male SD rats. Different concentrations of each of these hormones were prepared in the range of 0.01-60 ng/ml via serial dilution of each of the pure hormones using water for injection as a diluent. 30 µL of respective concentrations was then thoroughly mixed with 30 µL of blood and 150 µL of AgNPs. The major goal of adding these varied amounts to blood was to separately identify the GH and TE concentration-sensitive SERS bands in blood. Concentrations were then grouped as follows: low (0.01-2) ng/ml, medium (3-22) ng/ml, and high (25-60) ng/ml for TE; and low (0.01-2) ng/ml, medium (3-25) ng/ml, and high (30-60) ng/ml for GH. Blood was collected from SD rats and stored at the University of Nairobi's Department of Medical Physiology under the conditions outlined by Ondieki and the coauthors [27]. For SERS (Raman) measurements, spectral data were collected and recorded from 12 random spots on each sample using a portable Raman spectrometer (EZRaman-N Portable Analyzer System, Enwave Optronics, Inc.). The experimental parameters were: ~150 mW excitation power,

5 s exposure time, 5 s; spectral averaging, and boxcar, 1. Data pre-processing included background correction (Auto Baseline 2) and smoothing (box car of 1) using built-in EZRaman Analyzer software. The spectrometer spectral range was 100-4200 cm$^{-1}$ with 7 cm$^{-1}$ spectral resolution and was equipped with a 1.2 m HRP-8 high throughput high Rayleigh rejection fiber optics probe. The probe delivered the excitation beams and collected the Raman-scattered radiation to and from the sample. Raman spectral data analysis together with further preprocessing that included normalization to maximum intensity, principal component analysis (PCA), and analysis of variance (ANOVA) was done within the 'fingerprint' spectral region 600–1850 cm$^{-1}$ using a script in MATLAB R2021a (version 9.10.0.1602886, The MathWorks Inc., Natick, MA).

## 3. Results and Discussions

### *3.1. Characteristic SERS spectra of Pure Testosterone and growth hormones*

GH and TE are potent anabolic hormones that work together to improve growth and body composition [11]. Their structural formulae are $C_{23}H_{40}O_3$ ( GH) and $C_{39}H_{60}N_8O_{13}$ ( TE) [28]. The chemical structures of these hormones contain functional groups, such as carbonyl (C=O), carboxyl (COOH), and hydrocarbon (CH) groups; therefore, it was expected that some vibrations would be isoenergetic. Indeed, as shown in Fig. 1, common Raman bands (TE and GH) were observed at approximately 770, 948, and 1000-1400 cm$^{-1}$ which are attributable to vibrations in the hydrocarbon [29], carbonyl, and carboxylic [30] functional groups, respectively. These bands can be tentatively ascribed to C-C bending (770 cm$^{-1}$) [21], C-O and C-C stretching (800-1200 cm$^{-1}$) [9,27], and Amide III (1220-1340 cm$^{-1}$) [31,32] vibrations. Unique bands were observed around 668, 914, 1488, and 1652 cm$^{-1}$ for TE and 1440, 1518, and 1700 cm$^{-1}$ for GH. These bands can be tentatively assigned to C-O-C stretching (668 cm$^{-1}$) [21], CH$_2$/CH$_3$ scissoring (1440 cm$^{-1}$) [21], Amide II (1518 cm$^{-1}$) [33], and C=O stretching of Amide I (1650-1700 cm$^{-1}$) [32,34]. These latter TE and GH characteristic Raman bands may not necessarily be observed in blood because the different solvent environments influence their bond vibrational energies. Therefore, it was necessary to spike the blood drawn from SD rats with the two hormones and measure their Raman spectra.

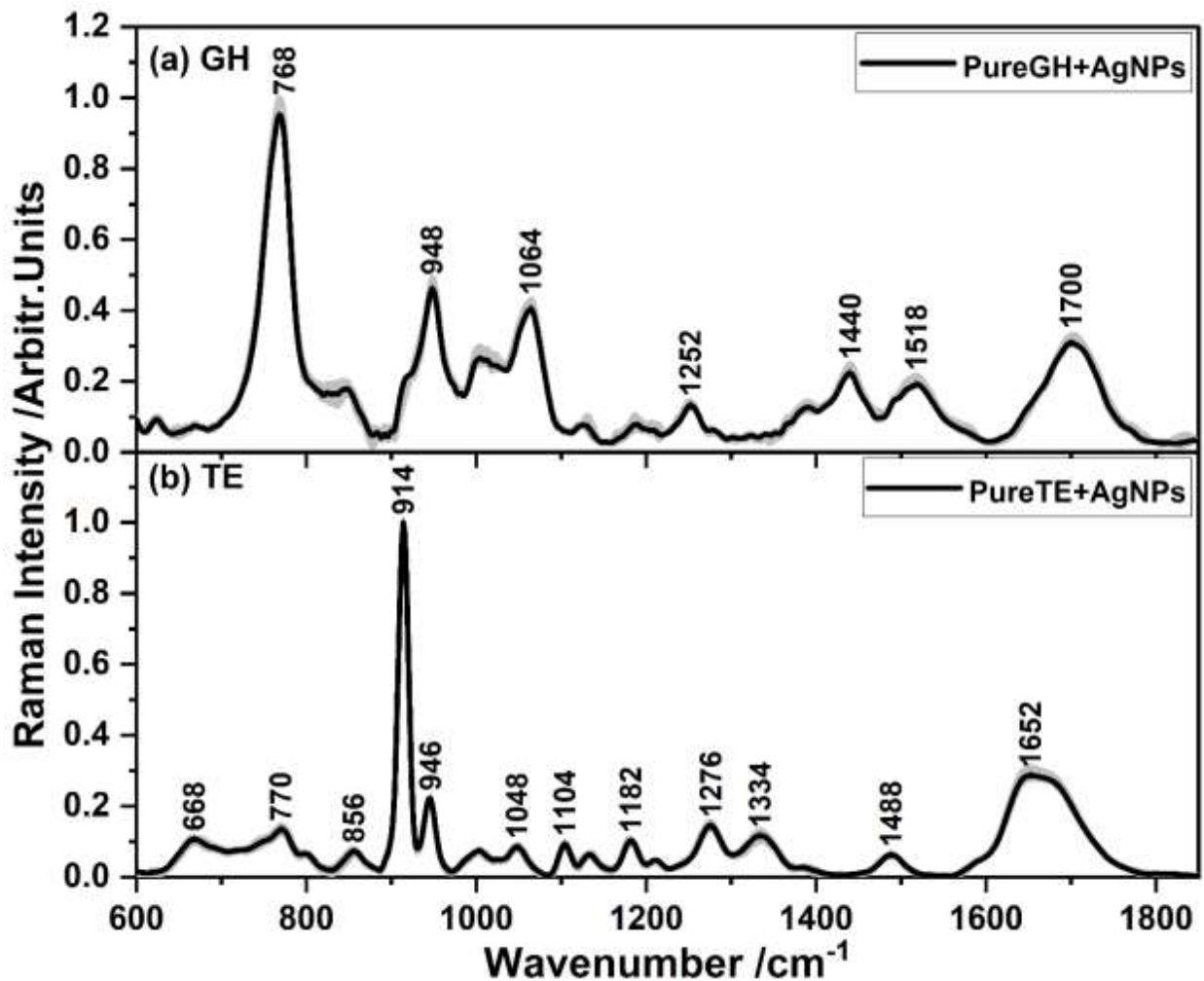

*Figure 1: Average SERS Spectra (black lines) derived from 12 spectra in each sample (grey lines) of (a) pure Growth Hormone and (b) pure testosterone hormones dissolved in water for injection.*

### 3.2. The SERS spectra of Testosterone and growth hormones when mixed with blood

Blood from male SD rats was spiked separately at various concentrations with TE and GH, and the Raman spectra were measured. Male rat blood was chosen because the levels of the two hormones (GH and TE) are known to fluctuate less [35]. Figure 2 shows the average SERS spectra for various hormone concentrations (GH and TE). As expected, the Raman spectral profiles from the hormone mixture in water (Fig. 1) and in the blood (Fig. 2) were dissimilar due to changes in the local environment around the molecules of the two hormones. The only similar bands (Fig 1 and Fig. 2) with slight shifts were those at approximately 1064 cm$^{-1}$ for GH, 910, 1190, 1490, and 1338 cm$^{-1}$ for TE, and 1658 cm$^{-1}$ for both hormones (TE and GH). Other bands, which are associated with various compounds in the blood, such as proteins, carbohydrates, and lipids, were observed at approximately 658, 798, and 1410 cm$^{-1}$.

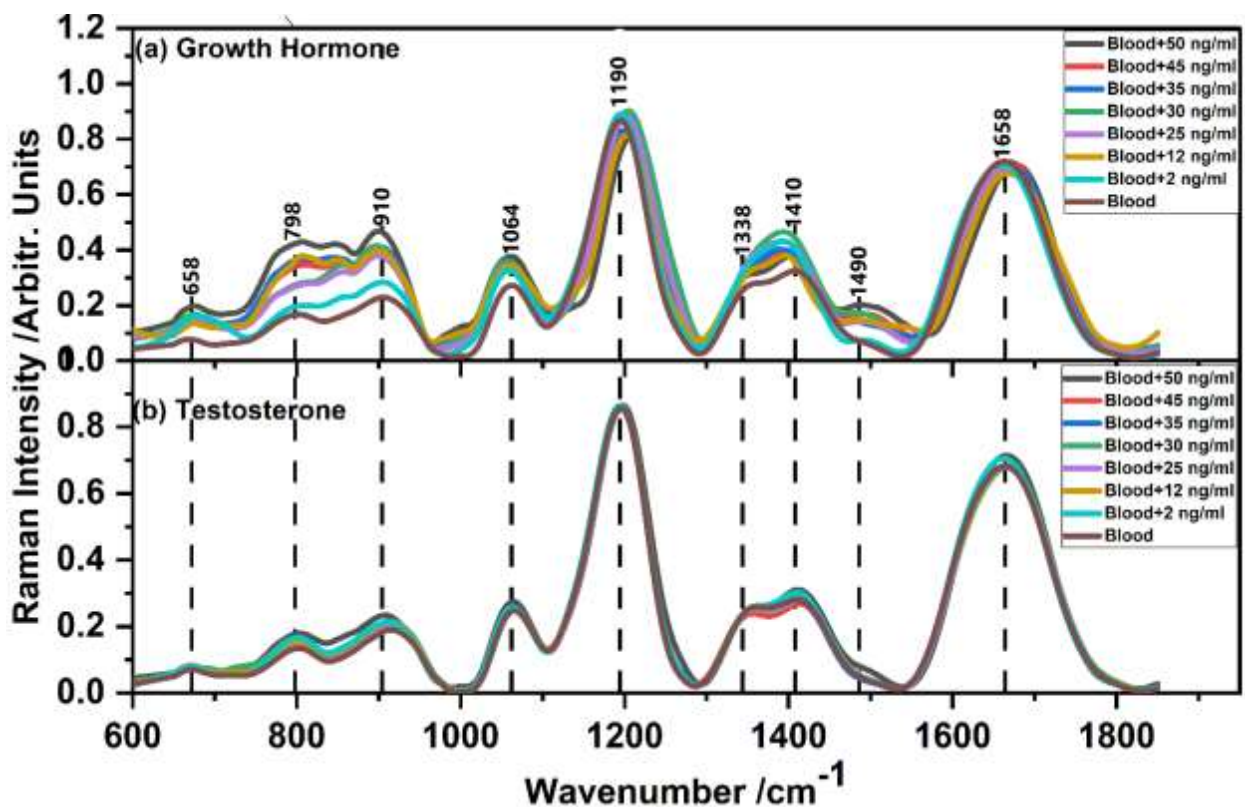

Figure 2: Mean SERS spectra derived from 12 spectra of each sample for different concentrations of (a) growth hormone, and (b) testosterone hormones in blood dried on a clean aluminum-wrapped glass slide.

These bands can be tentatively attributed to C-O-C stretching (658 cm$^{-1}$) [21], C-C bending (798 cm$^{-1}$) [21], C-O/C–C stretching (910 cm$^{-1}$) [9,32,36], C–C stretching (1064 cm$^{-1}$) [27], C–C stretching of β-carotene/tyrosine (1190 cm$^{-1}$), CH$_2$ wagging of proteins (1338 and 1410 cm$^{-1}$) [9,31], CH$_2$ bending (1490 cm$^{-1}$) [37], and carbonyl stretching of proteins (1658 cm$^{-1}$) [27,36].

To reveal subtle spectral differences between the two hormone samples in blood, principal component analysis (PCA) was performed on the combined data set. Segregation based on the spectral profile patterns is displayed in the score plot of Fig.3(a). The two spectral datasets (from GH and TE) were distinctly different. The bands responsible for segregation are displayed in the loading plots in Fig. 3 (b). From the loading plots, one can observe a subtle unique band centered around 1510 cm$^{-1}$ in GH, with the latter being overt in the spectra displayed in Fig. 2.

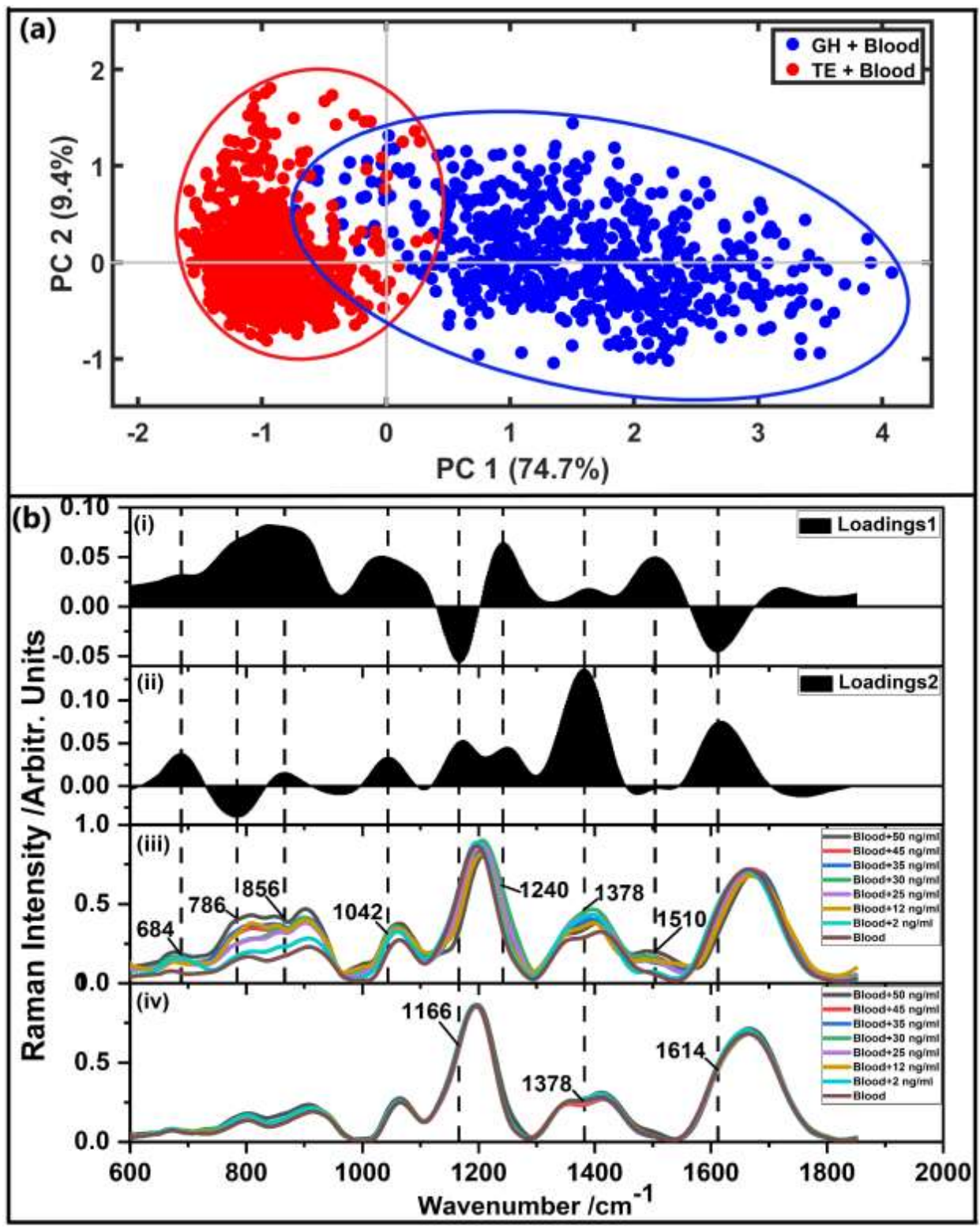

*Figure 3: (a) Two-dimensional PCA score plot for different hormone (GH and TE) concentrations in blood and (b) PCA loading plots ((i) and (ii)), Raman spectra of Growth hormone in the blood*

*(iii), and testosterone in the blood (iv). The explained variances are indicated in percentages, and were 74.7 and 9.4 % for PC1 and PC2, respectively.*

As also revealed by the PCA loadings plot (Figure 3(b)), the spectral segregation was brought about by several other bands that displayed large intensity variance. The bands that significantly contributed to segregation, as seen from the loading plot scores, were around 1378 cm$^{-1}$ (for GH and TE), 684, 786, 856, 1042, 1240, and 1510 cm$^{-1}$ (for GH), and 1166 and 1614 cm$^{-1}$ (for TE). The two PCs had a total explained variance of 84.1%, with PC1 and PC2 accounting for 74.7% and 9.4% of the variance, respectively. The remaining 15.9% of data not explained by the two PCs could be due to the components present in the blood such as haemoglobin (1378 cm$^{-1}$).

For the SERS method to be applied to detect or assay GH and TE separately, the Raman band intensities should vary with the concentration of each hormone. In this study, the Raman spectral data sets from various concentrations of GH and TE in blood were grouped as Low (0.01-2 ng/ml), Medium (3-22 ng/ml), and high (25-60 ng/ml) before performing PCA. PCA was performed to identify the bands (including subtle bands) that were sensitive to concentrations. Figure 4 shows the PCA scores and loading plots for the GH and TE datasets.

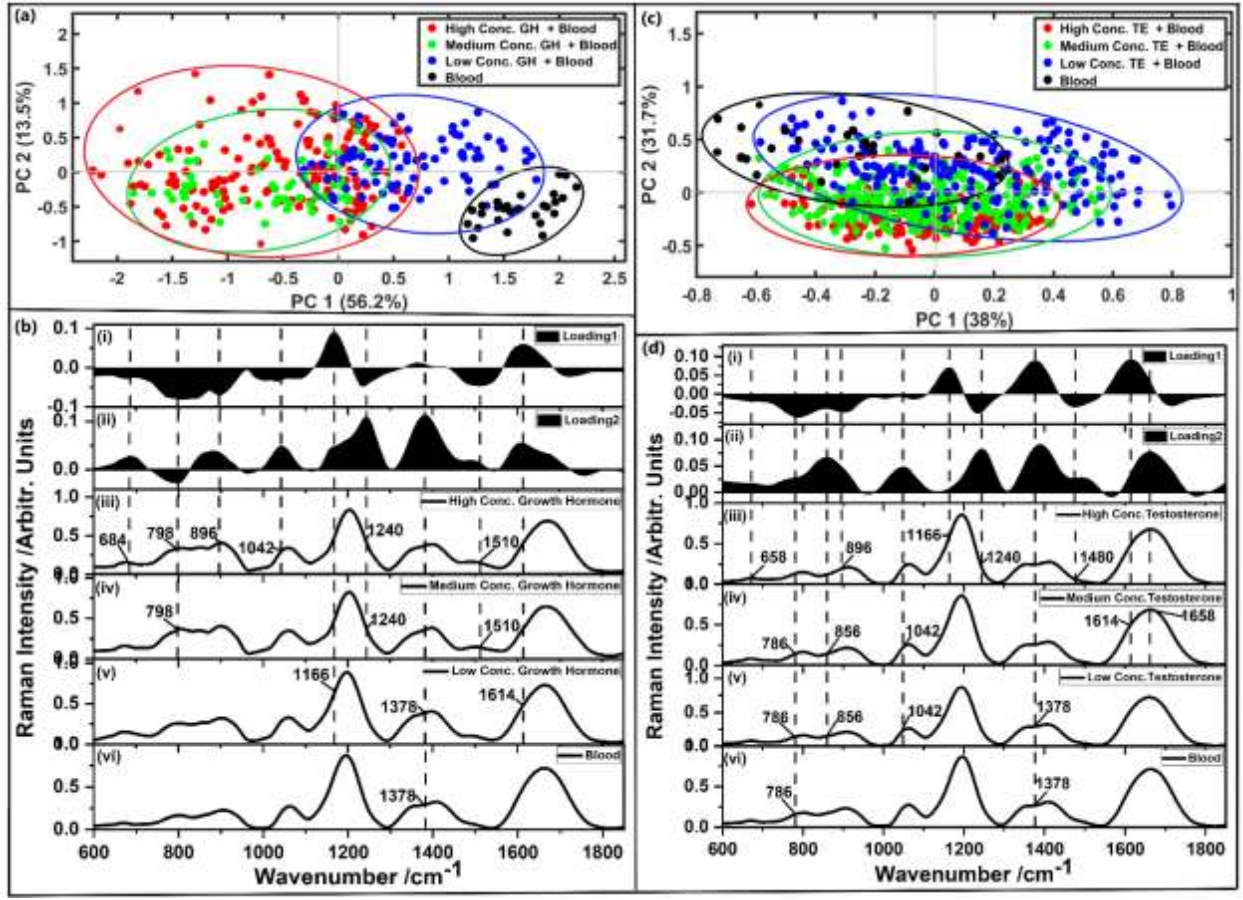

*Figure 4: The 2-Dimensional PCA score plots and loading plots for GH and TE mixed with blood at different concentrations. The explained variances are indicated in percentages and were 56.2 and 13.5 % for GH, and 38 and 31.7% for TE, for PC1 and PC2, respectively.*

In the Raman dataset from GH samples, it was evident from the score plot of Fig 4(a) that the spectra from non-spiked blood were segregated from the spiked ones. At the same time, low concentrations were distinctly segregated from the medium and high concentrations. The latter two (medium and high) were not significantly segregated. The bands responsible for the segregation were those centered at 1378 cm$^{-1}$ for both blood and blood with low concentrations of GH; 1166 and 1614 cm$^{-1}$ for low concentrations; 798, 1240, and 1510 cm$^{-1}$ for both medium and high concentrations; and 684, 896, and 1042 cm$^{-1}$ for high concentrations. For the Raman dataset of the TE samples, the groups (blood, low, medium, and high) were slightly segregated (Figure 4(c)). The bands that brought about the segregation were those centered around 786 cm$^{-1}$ for blood, low and medium concentrations of TE; 1378 cm$^{-1}$ for both blood and low concentrations of TE; 856 and 1042 cm$^{-1}$ for low and medium concentrations of TE; 1614 and 1658 cm$^{-1}$ for medium concentrations of TE; and 658, 896, 1166, 1240 and 1480 cm$^{-1}$ for high TE concentrations. As shown in Figures 4(b) and 4(d), some of the bands identified to bring about the differences in the spectral profiles of different concentrations of the two hormones matched exactly and some with slight variances from the ones in pure hormones (see figure 1). These bands include those around 798, 1042, and 1490 cm$^{-1}$ for GH and 658, 786, 896, 1042 1166, 1378, and 1658 cm$^{-1}$ for TE. A spectral band at 1378 cm$^{-1}$ was present for both hormones at low concentrations because of their components in the blood. The concentration-sensitive bands observed in the PCA results for the two hormones were centered at 896 and 1240 cm$^{-1}$ for all hormones; 684, 798, 1042 and 1510 cm$^{-1}$ for GH, and 658, 1166, and 1480 cm$^{-1}$ for TE.

Because both hormones (TE and GH) are generally present in blood at normal concentrations, it was important to investigate whether the concentration-sensitive bands in the GH and TE spectra also exist in a sample spiked with both at identical concentrations. As seen in the score plot displayed in Fig.5(a), there was some degree of segregation of the Raman spectral data sets from low, medium, and high concentrations, and non-spiked blood. The bands responsible for the segregation (with large and unique PC loadings) were those around 684 and 1614 cm$^{-1}$ in GH and 786, 856, and 1490 cm$^{-1}$ in TE. These bands qualify as biomarker bands for the respective hormones in the blood, as they are also noticeable in the TE and GH spectra displayed in Fig 4.

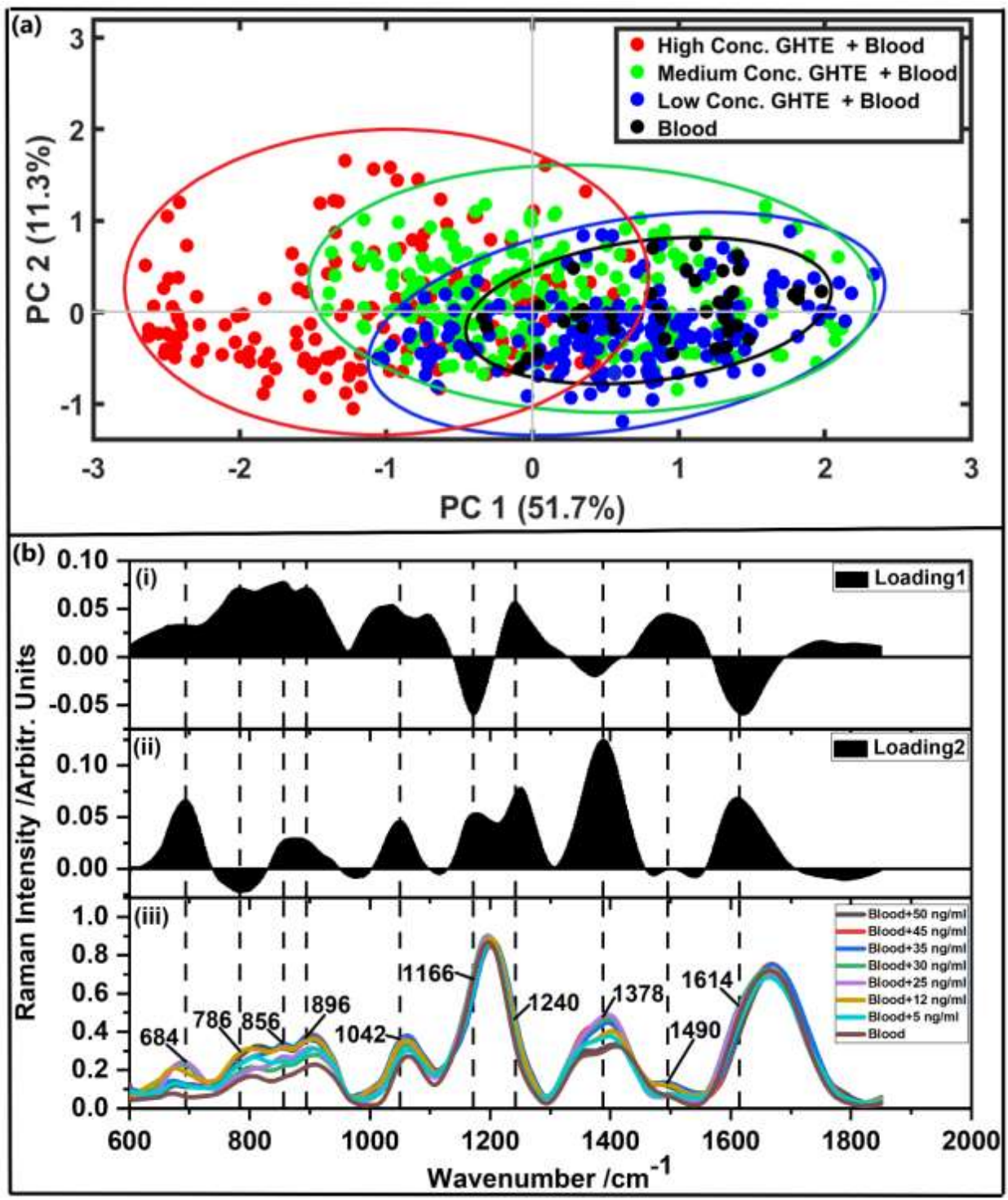

*Figure 5: (a) Two-dimensional PCA score plot, and (b) PCA loading plots ((i) and (ii)), and Raman Spectra of blood (iii), with both GH and TE mixed at different concentrations. The explained variances are indicated in percentages and were 51.7 and 11.3 % for PC1 and PC2, respectively.*

ANOVA was used to further confirm the concentration-sensitive bands obtained by PCA (Figure 6). It was expected that ANOVA would determine the specific spectral bands that showed significant differences across the groups (blood, low, medium, and high concentrations), and which specific group comparisons contributed to these differences. Thus, wavenumbers with significantly different means could serve as potential biomarkers or concentration-sensitive bands.

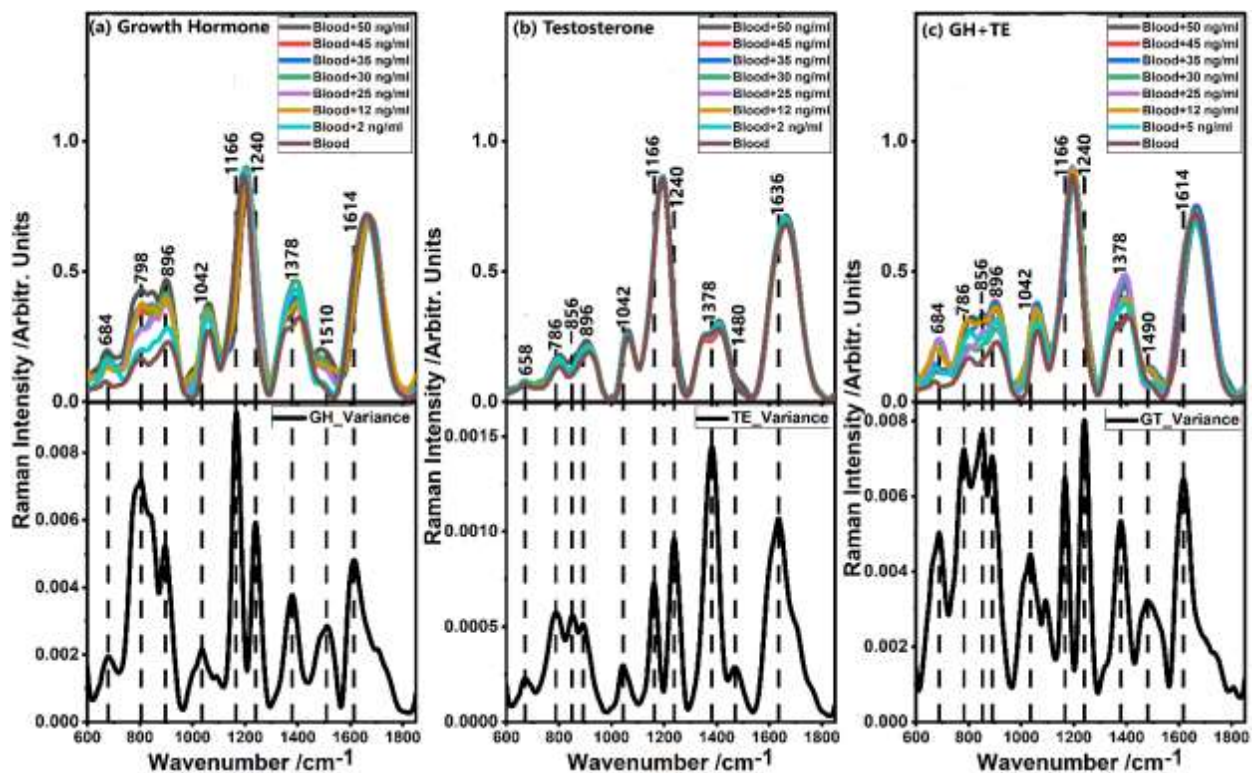

*Figure 6: Analysis of Variance results for SERS spectra of different concentrations of a) GH, b) TE, and c) both GH and TE mixed with male SD rat blood.*

As can be seen in Figure 2, the spectral profiles of different concentrations of GH and those of TE in the blood are fairly identical, with variations in the intensity of some bands (see Figure 5(a) and 5(b)). The profiles were identical since all the hormones investigated here are naturally present in blood at various concentration levels. ANOVA results for each hormone (Fig. 6(a) and 6(b)) and those from the combined (both GH and TE) hormones (Fig. 6(c)) showed prominent bands centered at wavenumbers around 684 cm$^{-1}$ (assigned to C-C Stretching) [38] and 1614 cm$^{-1}$ (assigned to C-C stretching) [38] in GH, and 786 cm$^{-1}$ (assigned to N-H wagging) [39], 856 cm$^{-1}$ (assigned to C-C stretching) [40], and 1490 cm$^{-1}$ (assigned to $CH_2$ bending) [37] in TE. This confirmed the use of these bands as concentration-sensitive bands for the two hormones. Other concentration-sensitive bands noted in each of the two hormones were around 896 cm$^{-1}$ (C-C Stretching) [36], 1042 (assigned to C-C stretching vibrations) [8,21], 1166 (assigned to C-O stretching and COH bending) [41], 1240 (assigned to Amide III) [36,41], and 1378 (C-C stretching) [9,31] (see Table 1). In addition, some variance was noted in bands centered at 1510 cm$^{-1}$ (attributable to CH$_2$ bending) [21,37] for GH and 1636 cm$^{-1}$ (assigned to Amide I) [4,32] for TE, which could be a potential biomarker band for GH and TE in the blood, respectively. The ANOVA results from the Raman spectral datasets from blood samples with different concentrations of both hormones showed a combination of ANOVA bands obtained in blood samples with each hormone. This implies that it is possible to identify the blood with both hormones. In addition, the consistency and overlap between the relevant spectral bands with high loadings on the important PCs discovered in PCA and the spectral regions identified as significant by ANOVA suggest that the spectral variations captured by PCA corresponded to those observed by ANOVA (Table 1).

A histogram plot of the combined variance of TE and GH was constructed to compare the intensity variation (from ANOVA) in similar bands (Fig. 7).

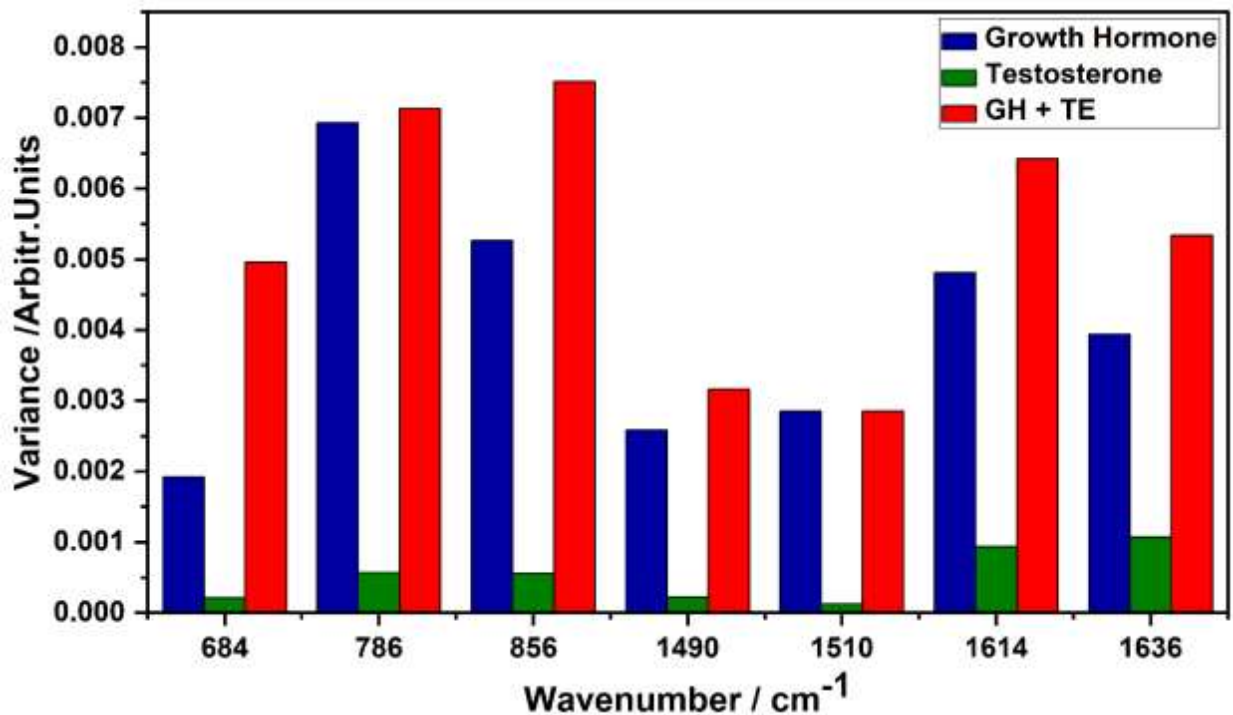

*Figure 7: Concentration-sensitive SERS band variation for each hormone.*

As shown in Figure 7, samples with a combination of the two hormones had the highest variance in all bands, followed by those with GH and TE. This suggests that the highest variance could be due to the combined variance of each of the two hormones in the blood.

### 3.3. Discussion

The simultaneous detection of GH and TE is important because these hormones can be abused singly or simultaneously [14]. The often-used methods such as UHPLC/HRMS and ELISA are not only limited to sensitivity, selectivity, and complex sample preparation but some, such as ELISA, can only assay one hormone at a time [19,20]. SERS, a technique that overcomes most of the limitations of traditional methods, has demonstrated the ability to detect these two hormones concurrently in the blood without interfering with each other. Here, SERS was noted to be highly sensitive for detecting different concentrations of GH and TE in the blood of SD rats. The prominent SERS bands centered at 658, 798, 910, 1064, 1382, 1490, and 1658 cm$^{-1}$, which demonstrated significant intensity variation with the amount of the two hormones in the blood, were of particular interest. In blood, the most concentration-sensitive bands (in both TE and GH), as deduced through Principal Component Analysis (PCA) and Analysis of Variance (ANOVA), were around 684 cm$^{-}$ and 1614 cm$^{-1}$ in GH; and 786, 856, and 1490 cm$^{-1}$ in TE. In addition, a characteristic variance was noted in the bands around 1510 cm$^{-1}$ (in GH) and 1636 cm$^{-1}$ (in TE), which could be potential biomarkers for the respective hormones in the blood and can be used to calibrate a SERS spectrometer to quantify their concentration in blood. Using a customized SERS system, it is possible to determine the levels of these hormones in the blood in less than a minute. This may also be used to distinguish between the high levels of these hormones owing to doping and physiology [42]. It should be made clear that no SERS system has been specialized to perform such tests, as many people are unaware of its existence and potential. Measurement of these hormone levels is crucial in sports because their use is prohibited and can lead to health risks [15,17].

### 4. Conclusion

This research demonstrated the potential use of the SERS technique for assaying GH and TE hormones in the blood. Concentration-sensitive SERS bands that exhibited prominent intensity variations with the concentrations of GH and TE were identified. These bands can be used to determine the levels of each hormone in the blood. The pure hormones displayed unique spectral profiles, thus demonstrating the potential quality checks for commercially available samples.

These findings widen the potential use of SERS in sports science, clinical diagnostics, and biomedical research. Here, a customized SERS system would be developed to assay GH and TE in blood where SERS metallic nanoparticles of the same size and shape should be used as SERS substrates for reproducibility purposes.

**Acknowledgments**

We sincerely express our gratitude to the Swedish International Development Cooperation Agency (SIDA) through the International Science Programme (ISP), Uppsala University, for sponsoring this research. We also acknowledge the Boniface Chege of the Department of Medical Physiology, University of Nairobi, for assisting with blood sample collection.

*Table 1: Raman bands obtained from PCA and ANOVA for hormones (GH and TE) in blood*

| **Wavenumber** | Growth hormone | | Testosterone | | Growth hormone + Testosterone | | Vibrational Assignment |
|---|---|---|---|---|---|---|---|
| | PCA | ANOVA | PCA | ANOVA | PCA | ANOVA | |
| 658 | | | √ | √ | | | C-O-C stretch [21] |
| 684 | √ | √ | | | √ | √ | C-C stretch [38] |
| 786 | | | √ | √ | √ | √ | N-H wagging [39] |
| 798 | √ | √ | | | | | C-C bending [21] |
| 856 | | | √ | √ | √ | √ | C-C Stretch [40] |
| 896 | √ | √ | √ | √ | √ | √ | C-C Stretch [36] |
| 1042 | √ | √ | √ | √ | √ | √ | C-C Stretch [8,21] |
| 1166 | √ | √ | √ | √ | √ | √ | C-O Stretch and COH bending [41] |
| 1240 | √ | √ | √ | √ | √ | √ | Amide III [36,41] |
| 1378 | √ | √ | √ | √ | √ | √ | Stretch (COO-) and C-H bend [8] |
| 1490 | | | √ | √ | √ | √ | $CH_2$ bending [37] |
| 1510 | √ | √ | | | | | Amide II [33] |
| 1614 | √ | √ | √ | | √ | √ | C=C stretch [32] |
| 1636 | | | | √ | | | Amide 1 [4,32] |
| 1658 | | | √ | | | | C-O/C=O Stretching [21,31] |